\documentclass[twocolumn,prl]{revtex4}
\usepackage{graphics,graphicx}
\begin{document}
\title{Multiple Magnetic Structures of Correlated Ce-ions in Intermetallic CeAu$_{2}$Ge$_{2}$}
\author{D.K.~Singh$^{1,2}$}
\author{A.~Thamizhavel$^{3}$}
\author{J.W.~Lynn$^{1}$}
\author{S.K.~Dhar$^{3}$}
\author{T.~Hermann$^{1}$}
\affiliation{$^{1}$NIST Center for Neutron Research, Gaithersburg, MD 20899, USA}
\affiliation{$^{2}$Department of Materials Science and Engineering, University of Maryland, College Park, MD 20742,USA}
\affiliation{$^{3}$Tata Institute of Fundamental Research, Mumbai, INDIA}

\begin{abstract}

Neutron diffraction measurements on a high quality single crystal of CeAu$_{2}$Ge$_{2}$ reveal two new magnetic structures that coexist with commensurate long range antiferromagnetic order below the Neel temperature $T_N$ $\simeq$14 K.  The new magnetic structures, which exhibit distinct temperature and field dependencies, are described by both commensurate and incommensurate magnetic structures and introduces new region in the $H$-$T$ phase diagram. These new experimental observations provide essential information for developing a universal understanding of the magnetic properties of this class of Ce-based Ce$X$$_{2}$$M$$_{2}$ (X = Cu, Ag, Au; M = Ge, Si) compounds, which are prototypes for exploring quantum phase transitions and their interplay with unconventional superconductivity.
\end{abstract}

 \maketitle

The study of Ce-based intermetallic compounds has proved to be a fertile testing ground for many important phenomena in strongly correlated electron systems, such as magnetic quantum critical phenomenon and its interplay with superconductivity, and non-Fermi liquid behavior.\cite{Gegenwart,Steglich} In recent years, a significant amount of research in this field has focused on understanding the underlying mechanism behind quantum phase transitions\cite{Hertz,Millis}-- as found in many Ce$X$$_{2}$$M$$_{2}$ compounds where $X$ is a transition metal element and $T$ is Si or Ge-- and develop a universal formulation which will have strong impact in understanding the physics of unconventional magnetic superconductors.\cite{Mathur,Coleman,Stewart,Lohneysen} Ce$X$$_{2}$$M$$_{2}$ crystallizes in the ThCr$_{2}$Si$_{2}$-type tetragonal structure ($I$4/$mmm$ space group), where the Ce valence spans the range from fully trivalent to strongly mixed valent as $X$ and $T$ are varied.\cite{Maple,Grier,Lawrence} Depending on the strength of the Ce ion $f$-band electron hybridization with the delocalized bands, the ground state properties can vary from the Kondo-interaction-dominated single-ion phenomenon to the Ruderman-Kittel-Kasuya-Yoshida (RKKY) interaction mediated long range magnetic ordered state.\cite{Si} The RKKY interaction acts as the mediating agent between the hybridized quasiparticles which often lead to the spin-density-wave instability of the local moments of $f$-electrons in the Fermi liquid of itinerant electrons.

In a subset of Ce$X$$_{2}$$M$$_{2}$ compounds where $T$ = Ge, neutron scattering measurements have revealed a spin-density wave type ground state magnetic configuration for $X$ = Cu and Ag.\cite{Loidl,Singh1,Singh2} Since CeAu$_{2}$Ge$_{2}$, which is the weakest heavy fermion metal of the Ce$X$$_{2}$Ge$_{2}$ series,\cite{Loidl} belongs to the same noble metal group of elements ($X$ = Cu, Ag, Au)- a natural corollary is expected but has never been observed. The magnetic susceptibility of CeAu$_{2}$Ge$_{2}$ exhibits strong anisotropy with an easy [001] direction perpendicular to the tetragonal plane.\cite{Joshi} Previous measurements on a powder of CeAu$_{2}$Ge$_{2}$ showed that the correlated Ce-ions develop a long range antiferromagnetic order below $T_N$$\simeq$15 K.\cite{Loidl} In this article, we report elastic neutron scattering measurements on a high quality single crystal sample of CeAu$_{2}$Ge$_{2}$. The crystals were synthesized by high temperature solution growth method using Bi flux and the quality was verified using X-ray measurements.\cite{Joshi} Our neutron scattering measurements reveal new magnetic structures, in addition to the long range magnetic order found in previous powder measurements, that evolve as a function of temperature and are strongly field dependent. The main result of this paper is the observation of incommensurate magnetic vectors, $\textbf{q}$, with the temperature dependent propagation wave vector $\epsilon$ = (0.055,0.055,0.055) at 4.5 K in reciprocal lattice units of 2${\pi}/a$, 2${\pi}/b$ and 2${\pi}/c$, respectively, with $a$ = $b$ = 4.39 $\AA$ and $c$ = 10.46 $\AA$ at 10 K in the tetragonal unit cell. The incommensurate magnetic peaks appear in a narrow temperature range of 2.5 K $\leq$ $T$ $\leq$ 8.5 K. Interestingly, the incommensurate wave-vector emerges in all three reciprocal directions. An additional magnetic phase is also found to exist with the propagation wave vector $\epsilon_L$ = (0,0,0.25) along the crystallographic direction $L$, which exhibits a different temperature dependence compared to both the commensurate antiferromagnetic order and the incommensurate order. 

\begin{figure}
\centering
\includegraphics[width=13.0cm]{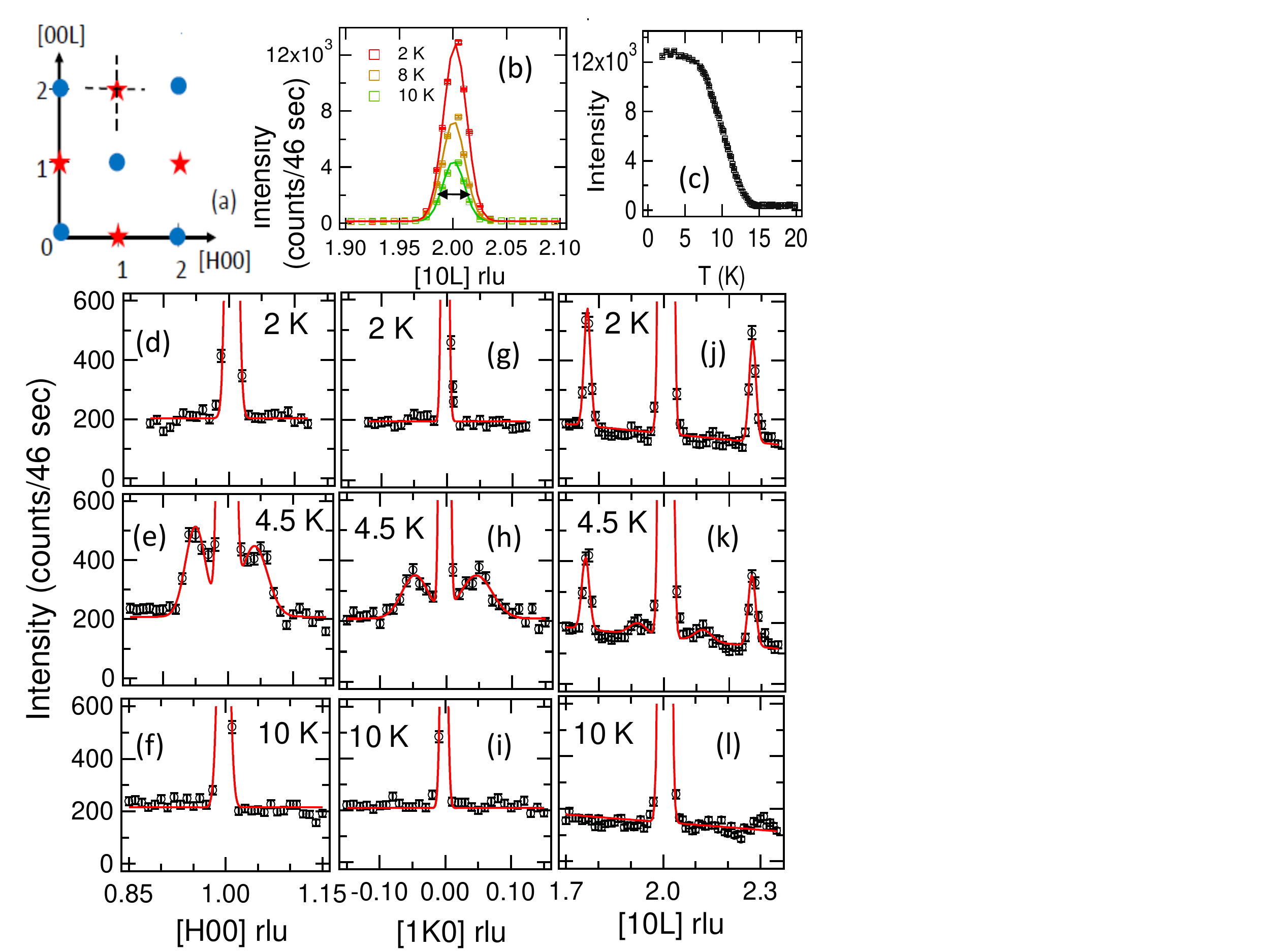} \vspace{-4mm}
\caption{(color online) (a) [H0L] scattering plane. Blue dots represent nuclear peaks and red stars represent magnetic peaks of the long range commensurate antiferromagnetic order (LRO). The black dashed lines are the elastic scan direction. (b) Representative scans at a few different temperatures across the [102] peak. Experimental data are well described by a Gaussian line shape. The double headed arrow, representing the resolution of the instrument, shows that the magnetic peaks are resolution limited. (c) Intensity of the [102] peak vs. temperature. The Neel$^{'}$ ordering temperature is $\simeq$14 K. (d)-(f) Elastic measurements along [H00] at a few different temperatures. As the temperature is changed from 2 K to 10 K, new incommensurate magnetic peaks appear, in addition to the commensurate LRO. A Gaussian line shape is found to be the best fit to the experimental data. (g)-(i) Similar behavior is observed in elastic scans along [1K0]. (j)-(l) In addition to the commensurate and incommensurate peaks described above, another set of commensurate magnetic peaks appear at L= 0.25, indicating a quadrupling of the magnetic unit cell along $c$-axis.. All three magnetic phases, LRO, incommensurate structure and the commensurate magnetic peaks along [10L] have distinct temperature dependences. Error bars in the article represent one standard deviation.
} \vspace{-4mm}
\end{figure}
 
Neutron scattering measurements were performed on a 0.7 gm single crystal sample of CeAu$_{2}$Ge$_{2}$ at the SPINS cold triple-axis spectrometer with fixed final neutron energy of 3.7 meV. The measurements employed a flat pyrolytic graphite (PG) analyzer with collimator sequence Mono-80$^{'}$-sample-BeO-80$^{'}$-flat analyzer-120$^{'}$. For higher wave vector resolution measurements at the fixed final neutron energy of 5 meV, the collimator sequence were Mono-40$^{'}$-sample-Be-40$^{'}$-flat analyzer-80$^{'}$. Measurements were performed with the crystal oriented in both the ($HK0$) and ($H0L$) scattering planes. For the measurements in magnetic field, the crystal was oriented in the ($HK0$) plane and mounted in a cryomagnet with a base temperature of $\simeq$2 K. For the magnetic field measurements, the field was applied along the easy [001] direction.

Previous measurements on powder CeAu$_{2}$Ge$_{2}$\cite{Loidl} identified magnetic reflections with commensurate long-range antiferromagnetic order.\cite{Loidl} Single-crystal measurements allow for a much more detailed examination of the intensities and magnetic scattering pattern which reveal qualiitatively new features. In Fig. 1b, we plot elastic representative scans at different temperatures across the [102] magnetic Bragg peak. As the sample is cooled below 14 K, resolution limited magnetic peaks clearly develop indicating the development of commensurate long range magnetic order in the system. A plot of peak intensity versus temperature at the magntic peak position [102] is shown in Fig. 1c. The low temperature elastic measurements show Bragg peaks at the positions $h+k+l = 2n+1$ (shown in Fig. 1a), which are systematically absent in the crystal structure. This shows that the ground state magnetic configuration is antiferromagentic in nature. The magnetic moment of the ordered Ce ion at 2 K is determined to be 1.65(5) $\mu_B$ with the moment along the c-axis.

\begin{figure}
\centering
\includegraphics[width=10.0cm]{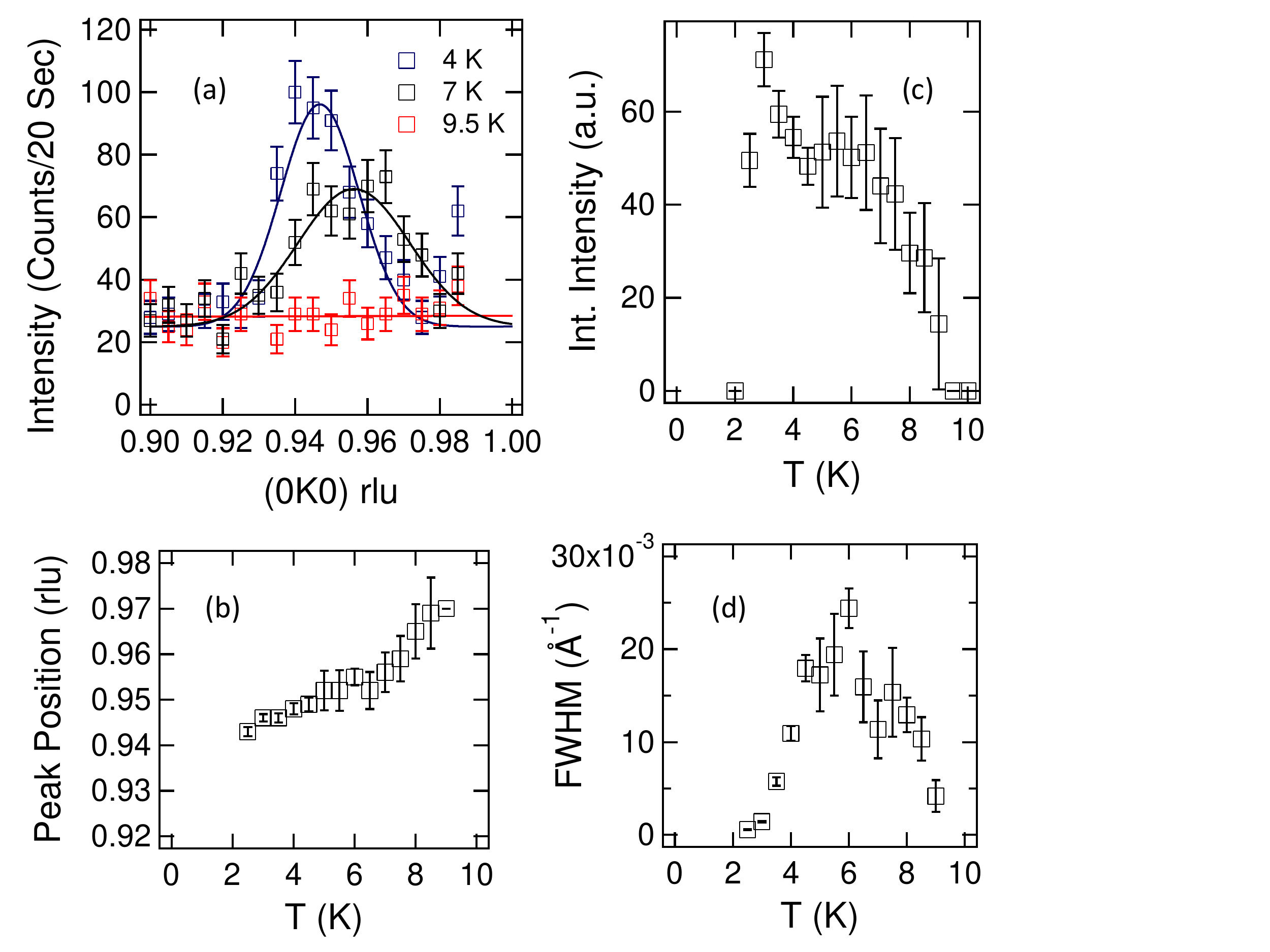} \vspace{-4mm}
\caption{(color online) (a) High resolution elastic scattering scans, with collimator sequence of M-40$^{'}$-S-Be Filter-40$^{'}$-A-80$^{'}$-D, along [0K0] across the incommensurate magnetic peak (0,0.945,0) at various temperatures. (b) The peak position of the incommensurate wave vector is found to vary with temperature. (c)-(d) Integrated intensity and net full width at half maximum as a function of temperature. Spectrometer resolution is subtracted from the full width at half maximum. The incommensurate peaks are broader than the instrument resolution at $T$$\simeq$9K. As the temperature decreases, the width of the peak becomes resolution limited at $T$$\rightarrow$2.5 K. 
} \vspace{-4mm}
\end{figure}

The observation of long range antiferromagnetic order at low temperature in CeAu$_{2}$Ge$_{2}$ is consistent with the previous measurements.\cite{Loidl,Joshi,Fritsch} However, as the sample is cooled below 10 K new magentic peaks start appearing with different temperature dependences. As shown in the series of plots in Fig. 1d-1l, two additional sets of magnetic peaks manifesting new magnetic components appear at $T$$\leq$10 K. In one case, incommensurate magnetic peaks of width broader than the instrument resolution develop across the long-range magnetic order peaks, for example [100], below 9 K. Most remarkably, the incommensurate peaks appear in all three crystallographic directions of [H00], [0K0] and [00L]. The modulation vector, $\epsilon$, descibing the incommensurate peak positions $\textbf{q}$, defined by $\textbf{q}$ =$\textbf{$\tau$}$$\pm$$\epsilon$ where $\textbf{$\tau$}$ and $\epsilon$ are the antiferromagnetic and modulation vectors respectively. $\epsilon$ is found to be temperature dependent with $\epsilon$ = (0.055, 0.055, 0.055) rlu at $T$ = 4.5 K. With further decrease in temperature $T$$\leq$2.5 K, the incommensurate peaks disappear. We performed a detail temperature dependence measurements of the incommensurate magnetic behavior at $\textbf{q}$ = (0,0.945,0) using the higher resolution collimator sequence. Representative scans at a few different temperatures at 2.5 K$\leq$$T$$\leq$9 K are plotted in Fig. 2a. A Gaussian line shape is found to best fit the experimental data. As the measurement temperature decreases, the position of the incommensurate peak clearly shifts away from the long-range order peak position at (010), Fig. 2b. In Fig. 2c and 2d, we have plotted the temperature dependencies of the integrated intensity and full width at half maximum (FWHM). The spectrometer resolution is subtracted from the fitted values of FWHM. As we can see in Fig. 2c, the integrated intensity of the incommensurate peak gradually rises below 9 K, attains a maximum at 3.5 K before suddenly disappearing below 2.5 K. At the same time, we find that the net width of the peak, which is slightly broader than the instrumental resolution at $T$$\simeq$8 K, becomes resolution limited at low temperature. We did not observe any lock-in transition or higher harmonics of the incommensurate wave vectors at $T$ = 2.5 K. We also note that the intensities of the incommensurate peaks are quite small, $\simeq$ 30 times, compared to the intensity of the long range order peaks. Because of the geometrical constraints at the spectrometer location, our measurements were limited to two antiferromagnetic positions [100] and [102] only. Therefore it was not possible to determine the nature of the magnetic configuration or, the ordered moment of Ce ions associated with the incommensurate magnetic structure. 

\begin{figure}
\centering
\includegraphics[width=12.0cm]{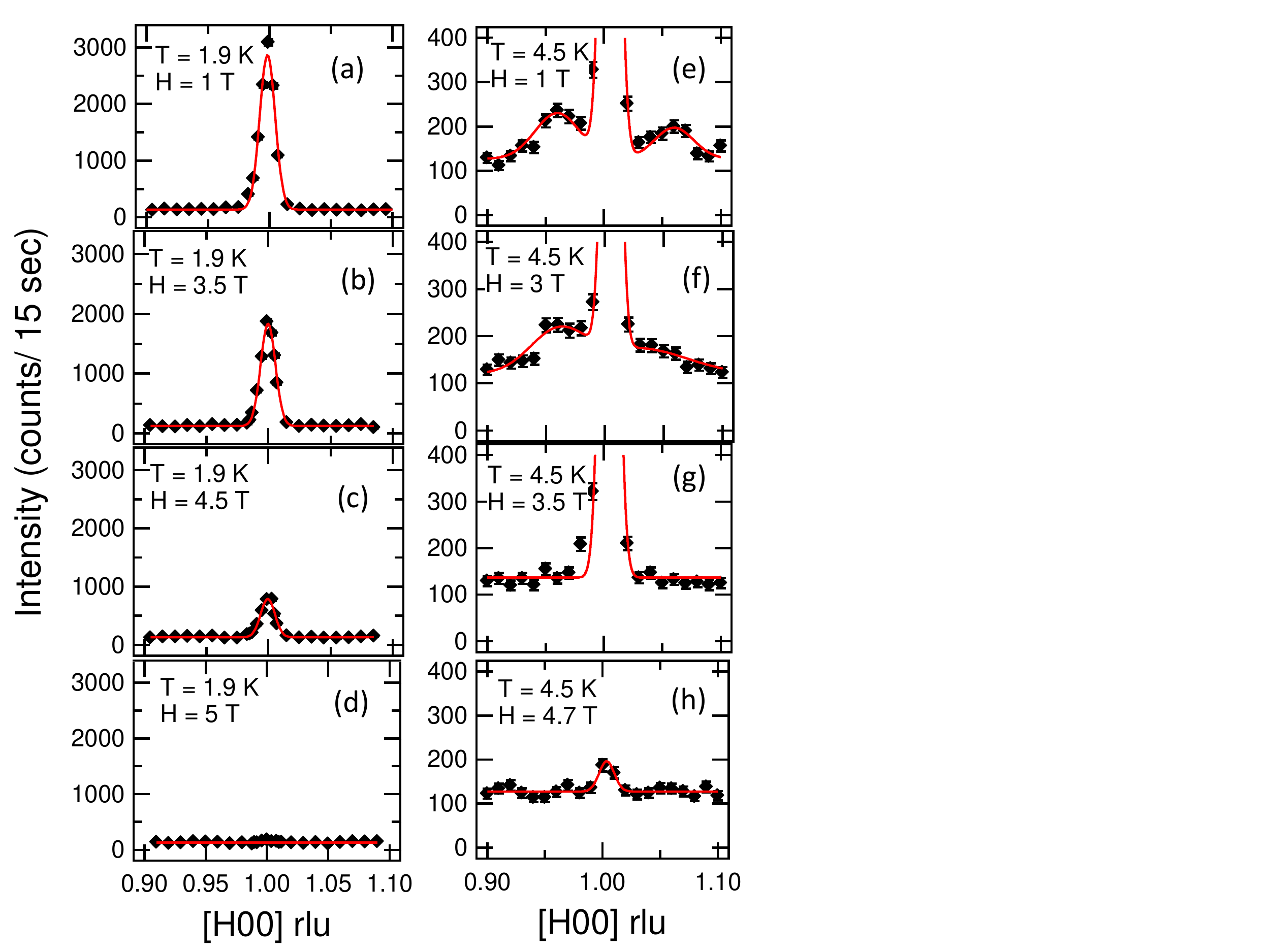} \vspace{-4mm}
\caption{(color online) Elastic scans along [H00] at various fields and two different temperatures. (a)- (d) At $T$ = 1.9 K, the long range order peak at [100] gradually disappears as a function of field. No incommensurate magnetic peaks exist at this temperature. (e)-(h) At $T$=4.5 K, the incommensurate (IC) magnetic structure appears in addition to the long range order (LRO). IC peaks exhibit a different field dependence than the LRO peaks. As the applied field is increased to $H$ = 3 T, the IC peaks become weaker. They disappear at $H$ $\simeq$ 3.5 T, while the LRO peaks disappears at $H$$\simeq$ 4.8 T at this temperature.
} \vspace{-4mm}
\end{figure}

In addition to the long range antiferromagnetic order and the incommensurate magnetic structure, a third type of magentic structure develops along the [00L] crystallographic direction. In this case, the resolution limited magnetic peaks appear below $T$$\simeq$~10 K and are described by propagation vector $\epsilon_L$ = (0,0,$\pm$0.25) rlu, as shown in Fig. 1j-1l. The magnetic peaks grow stronger as the temperature is reduced. A different temperature dependence of these magnetic peaks compared to both the long-range ordered AFM peaks and the incommensurate magnetic peaks suggests another new magnetic component of correlated Ce-ions in CeAu$_{2}$Ge$_{2}$. Since the third magnetic structure is represented by the commensurate magnetic peaks along the [00L] direction only, the magnetic configuration for this component of the Ce-moments can be described by an arrangement where the Ce spins have a component that points along either of the two equivalent crystallographic axes of [H00] or [0K0] and antiferromagnetically aligned with the nearest neighbors. Also the magnetic unit cell of such correlation is extended by four-times of the underlying lattice unit cell along the [00L] direction. All three magnetic components exhibit different temperature dependences and different types of magnetic correlations. 

Previous field-dependent work suggests that CeAu$_{2}$Ge$_{2}$ undergoes a metamagnetic transition at $H$$\simeq$4.5 T.\cite{Joshi} At sufficiently low temperatures, the nature of magnetic correlation of Ce ions changes from long-range antiferromagnet to field-induced ferromagnet at $H$$\geq$4.5 T. In Fig. 3, we have plotted the elastic scans at various fields at two temperatures, 1.9 K and 4.5 K. The measurements were performed for the crystal oriented in the ($HK0$) plane. As we can see, the peak intensity of the [100] peak, associated to the long range antiferromagentic order, gradually diminishes and eventually disappears at $H$ $\simeq$ 5.1 T at $T$ = 1.9 K. The field measurements at $T$ = 4.5 K reveal that the incommensurate peaks exhibit a different field dependence, compared to the primary long range antiferromagentic phase. In fact, the incommensurate peaks suddenly disappear at $H$ $\simeq$ 3.5 T, indicate a first order magnetic transition. The field dependence of this new phase is described in Fig. 4a. The distinct field-dependence provides additional evidence of multiple magnetic phases in CeAu$_{2}$Ge$_{2}$. We also tried to verify the possible existence of a field-induced ferromagentic phase at low temperature and high field. But we did not observe any such phase as the intensities of the nuclear peaks remained unchanged; Magnetic peaks associated with the ferromagnetic phase are expected to appear at the nuclear peak positions. Hence the intensities of the nuclear peaks are expected to be enhanced, which were not the case. 

\begin{figure}
\centering
\includegraphics[width=8.5cm]{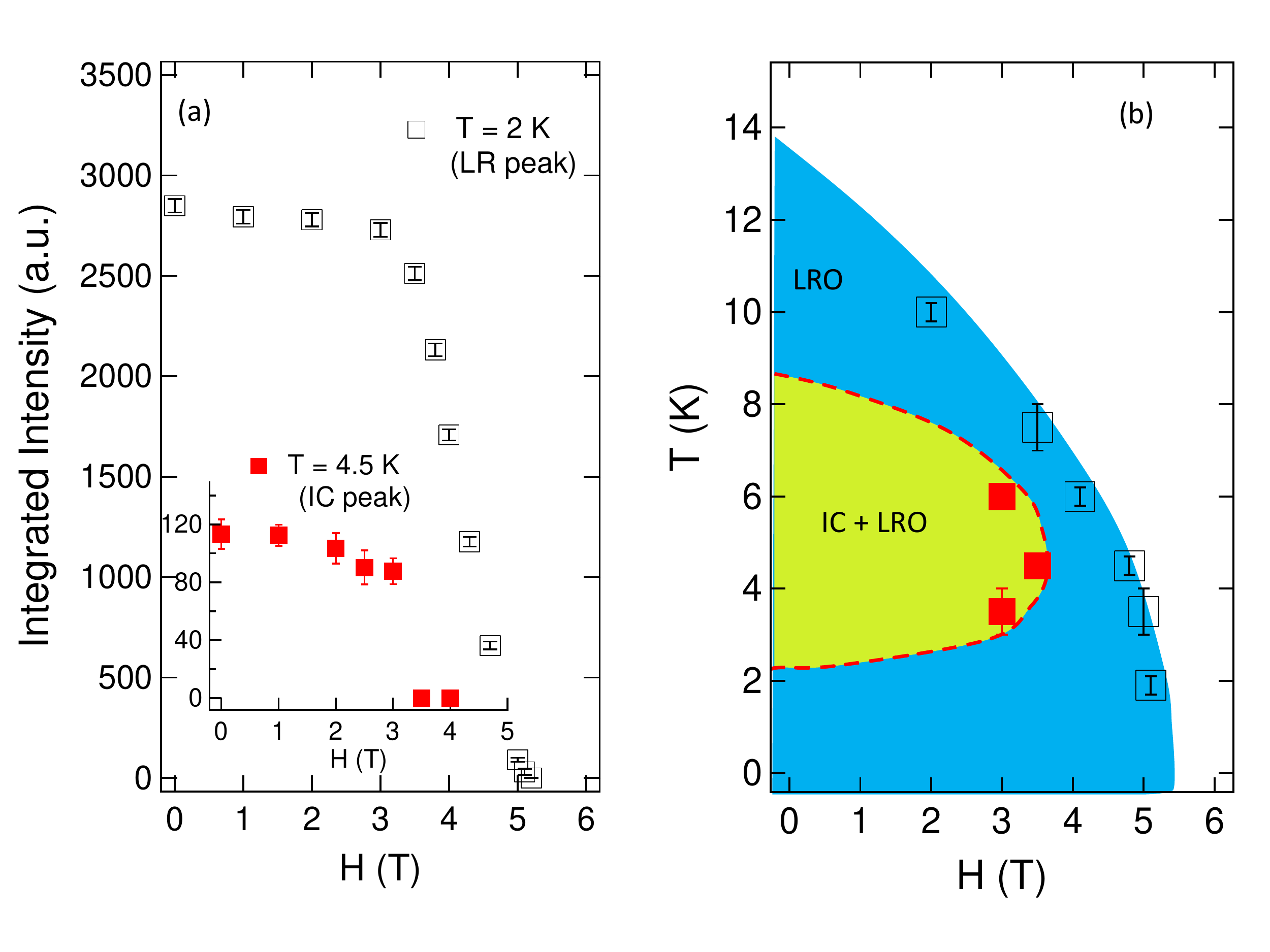} \vspace{-4mm}
\caption{(color online) (a) Integrated intensity vs. field for both LRO and a representative IC peak (inset) at $\textbf{q}$ = (0.955,0,0) (b) $H$-$T$ phase diagram for CeAu$_{2}$Ge$_{2}$. In addition to the long range order (LRO), the incommensurate structure (IC) also exists in a narrow $H$-$T$region.
} \vspace{-4mm}
\end{figure}

In Fig. 4b we have summarized our observations using an $H$-$T$ phase-diagram for elastic measurements in the [HK0] crystallographic plane. The additional new magnetic structures that develop below the primary ordering temperature have a different temperature and field dependencies than the quadrupolar order observed in PrPb$_{3}$ and therefore appear to have a different origin.\cite{Onimaru} A quadrupolar order generally results from the degenerate orbital level which carries a quadrupolar moment. These small moments are strongly field-dependent and become stronger as the applied field is increased. That is opposite to what we have observed. Another possibility is that these additional magnetic peaks could arise due to the intermixing of low-lying crystal field levels.\cite{Cox,Schlottmann} Previous as well as recent work on both polycrystalline and single crystal samples suggest that the crystal field level scheme in CeAu$_{2}$Ge$_{2}$ consists of three doublets at $\Delta_0$ = 0 K, $\Delta_1$ $\simeq$12 meV and $\Delta_2$ $\simeq$18 meV.\cite{Loidl,Joshi} The ground state is almost pure $J_z$ = $\pm$5/2 with an admixture of $\pm$3/2. The first excited doublet state is high enough in energy to avoid any significant intermixing between the ground and first excited state doublets, so that any induced moment magnetic state is highly unlikely. 

In order to understand the emergence of the new magnetic structures in CeAu$_{2}$Ge$_{2}$, we need to consider the origin and the type of magnetic order in the whole group of Ce$X_2$Ge$_{2}$ compounds, where $X$ = Cu, Ag and Au. In this group, both CeCu$_{2}$Ge$_{2}$ and CeAg$_{2}$Ge$_{2}$ compounds exhibit incommensurate spin-density wave type long range magnetic order. \cite{Singh1,Singh2} The spin-density wave type magnetic configuration originates from the strong hybridization between conduction bands and the 4$f$-band of Ce-ions via the Kondo interaction, which also gives rise to the heavy fermion character of the quasi-particles. It is interesting to note that as we move down the group, the heavy fermion character becomes weaker with CeCu$_{2}$Ge$_{2}$ as the heaviest.\cite{Loidl} Recent electrical transport measurements on single crystal CeAu$_{2}$Ge$_{2}$ reveal that the magnetoresistance (MR) initially decreases with field followed by an upward turn above $H$$\simeq$ 3 T and then decreases sharply above $H$$\simeq$4 T.\cite{Joshi} The authors explained their observation by invoking a weak Kondo interaction in this compound. The MR at $T$$\simeq$ 4  K is more negative than the MR at 2 K. Also, this behavior in MR is observed in a narrow temperature range of 2.5 K$\leq$$T$$\leq$8 K, which is the same temperature range where the weak incommensurate magnetic structure exists in CeAu$_{2}$Ge$_{2}$. Therefore, we argue that this new incommensurate phase in CeAu$_{2}$Ge$_{2}$ possibly results from the weak Kondo coupling between the 4$f$-band of Ce ions and the conduction bands. The weak heavy fermion character in this compound was also observed in the previous powder measurements.\cite{Loidl} Further experimental and theoretical investigations are highly desirable to understand in detail the new phase diagram proposed in Fig. 4. 

These new experimental results hint of the presence of a universal pattern in the magnetic configuration of this class of Ce-based materials. A comprehensive understanding of the magnetic states of Ce$X$$_{2}$$M$$_{2}$ compounds is highly desirable as many members of this class exhibit novel quantum critical phenomena and unconventional superconductivity. Future experimental work on the spin dynamics should prove particularly illuminating as soon as single crystal samples large enough for inelastic studies become available.

This work used facilities supported in part by the NSF under Agreement No. DMR-0944772.

\bibliography{CeAuGe}

\begin{thebibliography}{19}
\expandafter\ifx\csname natexlab\endcsname\relax\def\natexlab#1{#1}\fi
\expandafter\ifx\csname bibnamefont\endcsname\relax
  \def\bibnamefont#1{#1}\fi
\expandafter\ifx\csname bibfnamefont\endcsname\relax
  \def\bibfnamefont#1{#1}\fi
\expandafter\ifx\csname citenamefont\endcsname\relax
  \def\citenamefont#1{#1}\fi
\expandafter\ifx\csname url\endcsname\relax
  \def\url#1{\texttt{#1}}\fi
\expandafter\ifx\csname urlprefix\endcsname\relax\def\urlprefix{URL }\fi
\providecommand{\bibinfo}[2]{#2}
\providecommand{\eprint}[2][]{\url{#2}}

\bibitem[{\citenamefont{Gegenwart et~al.}(2008)\citenamefont{Gegenwart, Si, and
  Steglich}}]{Gegenwart}
\bibinfo{author}{\bibfnamefont{P.}~\bibnamefont{Gegenwart}},
  \bibinfo{author}{\bibfnamefont{Q.}~\bibnamefont{Si}}, \bibnamefont{and}
  \bibinfo{author}{\bibfnamefont{F.}~\bibnamefont{Steglich}},
  \bibinfo{journal}{Nature Physics} \textbf{\bibinfo{volume}{4}},
  \bibinfo{pages}{186} (\bibinfo{year}{2008}).

\bibitem[{\citenamefont{Steglich\emph{, et al.}}(1979)}]{Steglich}
\bibinfo{author}{\bibfnamefont{F.}~\bibnamefont{Steglich\emph{, et al.}}},
  \bibinfo{journal}{Phys. Rev. Lett.} \textbf{\bibinfo{volume}{43}},
  \bibinfo{pages}{1892} (\bibinfo{year}{1979}).

\bibitem[{\citenamefont{Hertz}(1976)}]{Hertz}
\bibinfo{author}{\bibfnamefont{J.~A.} \bibnamefont{Hertz}},
  \bibinfo{journal}{Phys. Rev. B} \textbf{\bibinfo{volume}{14}},
  \bibinfo{pages}{1165} (\bibinfo{year}{1976}).

\bibitem[{\citenamefont{Mills}(1993)}]{Millis}
\bibinfo{author}{\bibfnamefont{A.~J.} \bibnamefont{Mills}},
  \bibinfo{journal}{Phys. Rev. B} \textbf{\bibinfo{volume}{48}},
  \bibinfo{pages}{7183} (\bibinfo{year}{1993}).

\bibitem[{\citenamefont{Mathur\emph{, et al.}}(1998)}]{Mathur}
\bibinfo{author}{\bibfnamefont{N.~D.} \bibnamefont{Mathur\emph{, et al.}}},
  \bibinfo{journal}{Nature} \textbf{\bibinfo{volume}{394}}, \bibinfo{pages}{39}
  (\bibinfo{year}{1998}).

\bibitem[{\citenamefont{Coleman et~al.}(2001)\citenamefont{Coleman, Pepin, Si,
  and Ramazashvili}}]{Coleman}
\bibinfo{author}{\bibfnamefont{P.}~\bibnamefont{Coleman}},
  \bibinfo{author}{\bibfnamefont{C.}~\bibnamefont{Pepin}},
  \bibinfo{author}{\bibfnamefont{Q.}~\bibnamefont{Si}}, \bibnamefont{and}
  \bibinfo{author}{\bibfnamefont{R.}~\bibnamefont{Ramazashvili}},
  \bibinfo{journal}{J. Phys.: Cond. Matt.} \textbf{\bibinfo{volume}{13}},
  \bibinfo{pages}{R723} (\bibinfo{year}{2001}).

\bibitem[{\citenamefont{Stewarti}(2006)}]{Stewart}
\bibinfo{author}{\bibfnamefont{G.~R.} \bibnamefont{Stewarti}},
  \bibinfo{journal}{Rev. Mod. Phys.} \textbf{\bibinfo{volume}{78}},
  \bibinfo{pages}{743} (\bibinfo{year}{2006}).

\bibitem[{\citenamefont{Lohneysen et~al.}(2007)\citenamefont{Lohneysen, Rosch,
  , Vojta, and Wolfle}}]{Lohneysen}
\bibinfo{author}{\bibfnamefont{H.}~\bibnamefont{Lohneysen}},
  \bibinfo{author}{\bibfnamefont{A.}~\bibnamefont{Rosch}}, ,
  \bibinfo{author}{\bibfnamefont{M.}~\bibnamefont{Vojta}}, \bibnamefont{and}
  \bibinfo{author}{\bibfnamefont{P.}~\bibnamefont{Wolfle}},
  \bibinfo{journal}{Rev. Mod. Phys.} \textbf{\bibinfo{volume}{79}},
  \bibinfo{pages}{1015} (\bibinfo{year}{2007}).

\bibitem[{\citenamefont{Maple\emph{, et al.}}(2010)}]{Maple}
\bibinfo{author}{\bibfnamefont{M.~B.} \bibnamefont{Maple\emph{, et al.}}},
  \bibinfo{journal}{J. Low Temp. Phys.} \textbf{\bibinfo{volume}{161}},
  \bibinfo{pages}{4} (\bibinfo{year}{2010}).

\bibitem[{\citenamefont{Grier\emph{, et al.}}(1984)}]{Grier}
\bibinfo{author}{\bibfnamefont{B.~H.} \bibnamefont{Grier\emph{, et al.}}},
  \bibinfo{journal}{Phys. Rev. B} \textbf{\bibinfo{volume}{29}},
  \bibinfo{pages}{2664} (\bibinfo{year}{1984}).

\bibitem[{\citenamefont{Lawrence\emph{, et al.}}(1981)}]{Lawrence}
\bibinfo{author}{\bibfnamefont{J.~M.} \bibnamefont{Lawrence\emph{, et al.}}},
  \bibinfo{journal}{Rep. prog. Phys.} \textbf{\bibinfo{volume}{44}},
  \bibinfo{pages}{1} (\bibinfo{year}{1981}).

\bibitem[{\citenamefont{Si et~al.}(2001)\citenamefont{Si, Rabello, Ingersent,
  and Smith}}]{Si}
\bibinfo{author}{\bibfnamefont{Q.}~\bibnamefont{Si}},
  \bibinfo{author}{\bibfnamefont{S.}~\bibnamefont{Rabello}},
  \bibinfo{author}{\bibfnamefont{K.}~\bibnamefont{Ingersent}},
  \bibnamefont{and} \bibinfo{author}{\bibfnamefont{J.~L.} \bibnamefont{Smith}},
  \bibinfo{journal}{Nature} \textbf{\bibinfo{volume}{413}},
  \bibinfo{pages}{804} (\bibinfo{year}{2001}).

\bibitem[{\citenamefont{Loidl\emph{, et al.}}(1992)}]{Loidl}
\bibinfo{author}{\bibfnamefont{A.}~\bibnamefont{Loidl\emph{, et al.}}},
  \bibinfo{journal}{Phys. Rev. B} \textbf{\bibinfo{volume}{46}},
  \bibinfo{pages}{9341} (\bibinfo{year}{1992}).

\bibitem[{\citenamefont{Singh\emph{, et al.}}(2011{\natexlab{a}})}]{Singh1}
\bibinfo{author}{\bibfnamefont{D.~K.} \bibnamefont{Singh\emph{, et al.}}},
  \bibinfo{journal}{Phys. Rev. B} \textbf{\bibinfo{volume}{84}},
  \bibinfo{pages}{052401} (\bibinfo{year}{2011}{\natexlab{a}}).

\bibitem[{\citenamefont{Singh\emph{, et al.}}(2011{\natexlab{b}})}]{Singh2}
\bibinfo{author}{\bibfnamefont{D.~K.} \bibnamefont{Singh\emph{, et al.}}},
  \bibinfo{journal}{Scientific Reports} \textbf{\bibinfo{volume}{1}},
  \bibinfo{pages}{117} (\bibinfo{year}{2011}{\natexlab{b}}).

\bibitem[{\citenamefont{Joshi\emph{, et al.}}(2010)}]{Joshi}
\bibinfo{author}{\bibfnamefont{D.~A.} \bibnamefont{Joshi\emph{, et al.}}},
  \bibinfo{journal}{J. Mag. Mag. Mat.} \textbf{\bibinfo{volume}{322}},
  \bibinfo{pages}{3363} (\bibinfo{year}{2010}).

\bibitem[{\citenamefont{Onimaru et~al.}(2005)\citenamefont{Onimaru, Sakakibara,
  Aso, Yoshizawa, Suzuki, and Takeuchi}}]{Onimaru}
\bibinfo{author}{\bibfnamefont{T.}~\bibnamefont{Onimaru}},
  \bibinfo{author}{\bibfnamefont{T.}~\bibnamefont{Sakakibara}},
  \bibinfo{author}{\bibfnamefont{N.}~\bibnamefont{Aso}},
  \bibinfo{author}{\bibfnamefont{H.}~\bibnamefont{Yoshizawa}},
  \bibinfo{author}{\bibfnamefont{H.~S.} \bibnamefont{Suzuki}},
  \bibnamefont{and} \bibinfo{author}{\bibfnamefont{T.}~\bibnamefont{Takeuchi}},
  \bibinfo{journal}{Phys. Rev. Lett.} \textbf{\bibinfo{volume}{94}},
  \bibinfo{pages}{197201} (\bibinfo{year}{2005}).

\bibitem[{\citenamefont{Cox}(1987)}]{Cox}
\bibinfo{author}{\bibfnamefont{D.~L.} \bibnamefont{Cox}},
  \bibinfo{journal}{Phys. Rev. Lett.} \textbf{\bibinfo{volume}{59}},
  \bibinfo{pages}{1240} (\bibinfo{year}{1987}).

\bibitem[{\citenamefont{Schlottmann}(1984)}]{Schlottmann}
\bibinfo{author}{\bibfnamefont{P.}~\bibnamefont{Schlottmann}},
  \bibinfo{journal}{Phys. Rev. B} \textbf{\bibinfo{volume}{30}},
  \bibinfo{pages}{1454} (\bibinfo{year}{1984}).

\end{thebibliography}
\end{document}